\documentclass[showpacs,preprintnumbers,amsmath,amssymb,floatfix,prd,11pt]{revtex4}
\usepackage[utf8]{inputenc}
\usepackage{epsfig}
\usepackage{amsmath}
\usepackage[dvipsnames]{xcolor}

\begin{document}
\title{Dijet Production in Neutral Current and Charged Current Polarized Deep Inelastic Scattering}
\author{Ignacio Borsa}  
\email{iborsa@df.uba.ar}
\affiliation{Departamento de F\'{\i}sica and IFIBA, Facultad de Ciencias Exactas y Naturales, Universidad de Buenos Aires, Ciudad Universitaria, Pabell\'on\ 1 (1428) Buenos Aires, Argentina}
\author{Daniel de Florian}  
\email{deflo@unsam.edu.ar}
\author{Iv\'an Pedron}  
\email{ipedron@unsam.edu.ar}
\affiliation{International Center for Advanced Studies (ICAS), ICIFI and ECyT-UNSAM, 25 de Mayo y Francia, (1650) Buenos Aires, Argentina}

\begin{abstract}

We calculate the fully differential Next-to-Leading Order (NLO) corrections to (longitudinally) polarized dijet production in deep inelastic scattering, featuring both neutral current and charged current processes. We also analyze their phenomenological impact in the Electron-Ion Collider (EIC) kinematics.

\end{abstract}

\maketitle

\section{Introduction}

The way in which the proton's spin emerges from the share of its constituents is still an open question and a key area of research in particle physics. The amount of spin carried by quarks and gluons is codified in terms of polarized parton distributions functions (pPDFs), which can be probed in high energy collision processes with longitudinally polarized nucleons. In spite of the significant progress achieved in the determination of pPDFs in the last twenty years, our knowledge on these distributions remains limited, specially compared to that of their unpolarized counterparts. This difference is mainly due to the lack of a large amount of complementary measurements of observables probing a wide kinematical range, associated with the technical challenges involved in the production of polarized beams. 

Although the amount of spin carried by quarks and antiquarks has been confirmed to be lower than the naive-parton-model value of $\hbar/2$ by fixed-target polarized deep inelastic scattering measurements at CERN, SLAC, DESY, and JLAB~\cite{Aidala:2012mv}, it is still unknown how much of the remaining spin is carried by gluons or is associated with the orbital angular momentum of partons. Stronger constraints on the gluon distributions were obtained from measurements carried out by the RHIC spin program~\cite{Aschenauer:2013woa}, albeit for a reduced range in the proton momentum fraction $x$. In addition to the question regarding the separation of the quarks and gluon spin, the way in which each of the different flavors of quarks contributes to the proton spin is also work in progress. The determination of flavor-discriminated parton distribution is primarily driven by semi-inclusive deep inelastic scattering (SIDIS) data, which require previous knowledge on the parton-to-hadron fragmentation functions, as well as charged weak vector boson production in proton-proton collision data, again covering only a rather limited kinematical range. 

In that sense, the construction of the future Electron-Ion-Collider (EIC), with a wider coverage in both the proton momentum fraction $x$ and the boson virtuality $Q^2$, and reaching an unprecedented precision for polarized measurements~\cite{Accardi:2012qut}, is likely to shed some light on these questions and provide new insights on the spin structure of the proton~\cite{Aschenauer:2012ve,Aschenauer:2015ata,Aschenauer:2020pdk,Boughezal:2018azh}. Besides extending our knowledge on pPDFs towards lower values of $x$ and providing additional constraints on the gluon distribution through scaling violations, measurements of polarized DIS mediated by electroweak bosons at the EIC will be particularly important to discriminate the helicity associated with each of the different quark flavors. Electron-proton scattering processes receive contributions from the exchange of virtual electroweak bosons $Z$ or $W^{\pm}$, which become significantly relevant at higher values of the virtual boson virtuality $Q^2$. Both charged current (CC) and neutral current (NC) DIS offer crucial complementary information on the nucleon spin structure, since they probe different partonic combinations than their purely photonic counterpart. While the use of CC DIS data has become a standard tool to improve flavor separation in modern unpolarized PDF extractions, there is currently no data on CC DIS taken on longitudinally polarized nucleons. Thus, new measurements of this process at the EIC would provide a valuable addition to the corpus of existing data, providing new independent constraints to polarized PDFs~\cite{Aschenauer:2013iia}. Additionally,  $Z$ exchange in NC processes has the advantage of being accessible even at lower values of $Q^2$ due to the $\gamma Z$-interference with the photon in DIS. This means that the NC measurements at the EIC could provide new electroweak precision tests, with accurate determinations of the electroweak vector and axial-vector couplings, as well as be a probe for beyond-standard-model physics~\cite{Zhao:2016rfu, AbdulKhalek:2021gbh}.

New high precision measurements should be accordingly accompanied by high precision theoretical calculations of the corresponding observables. Even though significant advances were made during the past 30 years in the computation of higher order corrections for unpolarized processes, setting next-to-next-to-leading order (NNLO) as the standard for Large-Hadron-Collider (LHC) calculations and even reaching the following order in some cases, the picture for polarized calculations is not yet as developed. Until recently, NNLO QCD corrections for polarized processes were only known for completely inclusive Drell-Yan~\cite{Ravindran:2003gi} and DIS~\cite{Zijlstra:1993sh}, in addition to the helicity splitting functions~\cite{Vogt:2008yw,Moch:2014sna,Moch:2015usa}. Within the past year, the first polarized NNLO exclusive cross sections were obtained both for jet production in DIS~\cite{Borsa:2020ulb,Borsa:2020yxh} and $W$ boson production in proton-proton collisions~\cite{Boughezal:2021wjw}.

Following our previous work, the first step to reach NNLO accuracy for jet production in electroweak DIS lies in the calculation of the NLO dijet cross section. In this paper we extend the results in~\cite{Borsa:2020ulb,Borsa:2020yxh} and present the NLO calculation for dijet production in polarized lepton-nucleon scattering, including the contributions stemming from the exchange of a virtual $Z$ or $W$ boson. The calculation is based on our extension of the Catani-Seymour dipole subtraction method~\cite{Catani:1996vz} to account for polarized initial-state particles. The computation is fully differential in the jet and lepton momenta. We analyze the impact of higher order corrections in the EIC kinematics, its perturbative stability and phenomenological implications, for both neutral and charged current processes.

The rest of the paper is organized as follows: in section \ref{sec:dijetkin} we begin by defining the kinematics for dijet production in DIS. Then, in section \ref{sec:higher_order_corrections} we go into some of the intricacies of the calculation of polarized electroweak processes within dimensional regularization. In section \ref{sec:dijets} we present the phenomenological results for NLO inclusive dijet production at the EIC in the Breit-frame for both NC and CC processes. Finally, in section \ref{sec:conclusion} we summarize our work and present our conclusions.

\section{Dijet production kinematics}\label{sec:kinematics}
\label{sec:dijetkin}

We start by briefly reviewing the case of leading dijet production in DIS. We study the process $$l(k)+P(p)\rightarrow l'(k')+\mathrm{jet}(p_{T,1},\eta_{1})+\mathrm{jet}(p_{T,2},\eta_{2})+X,$$ where $k$ and $p$ are the momenta of the incoming lepton and proton, respectively, and $k'$ is the momentum of the outgoing lepton. We consider both neutral and charged current processes, mediated by the exchange of a virtual boson with momentum $q=k-k'$ and virtuality $Q^{2}=-q^{2}$ fully determined by the lepton kinematics. The Bjorken variable $x$ and the inelasticity $y$ are then defined as usual by 

\begin{equation}
    x=\frac{Q^{2}}{2 p\cdot q}, \qquad y=\frac{q\cdot p}{k\cdot p}.
\end{equation}

\noindent For CC electron-proton scattering it should be noted that, while the kinematics of the outgoing neutrino are not experimentally accessible, the values of $x$ and $Q^2$ can be reconstructed from the hadronic final state using the Jacquet-Blondel method. 

In addition to these variables, which are used in the analysis of fully inclusive DIS, a deeper insight on the underlying partonic kinematics can be obtained through the study of the final-state jets, which can be characterized in terms of their transverse momentum $p_{T,i}$ with respect to the beam, and its pseudorapidity $\eta_{i}$. The availability of two jets allows for an even more in-depth study of the partonic kinematics. As in the H1 \cite{Andreev:2014wwa,Andreev:2016tgi} and ZEUS \cite{Abramowicz:2010cka} experiments, and in addition to the jets' transverse momentum and pseudorapidities, the dijet production cross section can be studied in terms of variables such as the invariant mass $M_{12}$ of the two-jet system, the dijet momentum fraction $\xi_2$, as well as the pseudorapidity difference $\eta^{*}$ in the Breit-Frame, which are defined by

\begin{equation}\label{eq:di-jet_kinematics}
\begin{split}
    M_{12}&=\sqrt{(p_{1}+p_{2})^{2}},\\
    \xi_{2}&=x\big(1+\frac{M^{2}_{12}}{Q^{2}}\big),\\
    \eta^{*}&=\frac{1}{2}|\eta_{1}^B-\eta_{2}^B|.
\end{split}    
\end{equation}

\noindent It is worth noticing that, at the LO of dijet production, $\xi_{2}$ is the momentum fraction carried by the incoming parton.

\begin{figure}[t]
 \epsfig{figure= 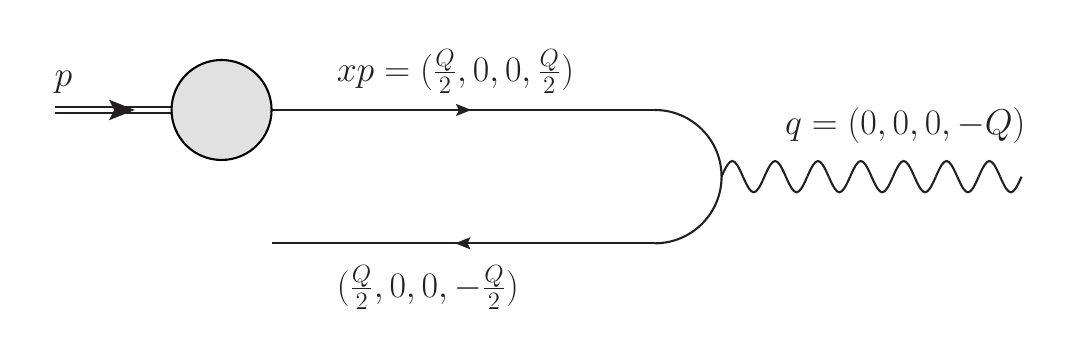 ,width=0.95\textwidth}
  \caption{The $\mathcal{O}(\alpha_{S}^{0})$ Breit frame kinematics for the process $p_{1}(xp)+\gamma^{*}(q)\rightarrow p_{2}(xp+q)$.}\label{fig:Breit}
\end{figure}

Dijet production can be better studied in the Breit frame (B), which is defined as the one that satisfies $2x\vec{p}+\vec{q}=0$. In the $\mathcal{O}(\alpha_{s}^{0})$ DIS process, this implies that the virtual boson and incoming parton collide head-on, completely reversing the momentum of the parton (hence the commonly used nickname \textit{brick-wall} frame), as represented schematically in Fig. \ref{fig:Breit}. The first nonvanishing contribution to dijet production is then obtained at $\mathcal{O}(\alpha_{s})$, with two final-state partons with completely opposite transverse momentum, and it receives contributions from both initial-state quarks and gluons.

\section{Calculation of higher order corrections}\label{sec:higher_order_corrections}

The calculation of cross sections beyond lowest order in QCD necessarily involves cancellations between individually divergent pieces stemming from infrared real emission and virtual diagrams, in addition to the factorization contributions, which can be archived via various subtraction or phase space slicing methods \cite{Catani:2007vq,GehrmannDeRidder:2005cm,Boughezal:2011jf,Cacciari:2015jma,Czakon:2010td,Binoth:2004jv,Anastasiou:2003gr,Somogyi:2006da,Stewart_2010}. In our code {\tt POLDIS}, the NLO dijet cross section is obtained by the implementation of the Catani-Seymour dipole subtraction \cite{Catani:1996vz}, extended to account for initial-state polarized particles \cite{Borsa:2020ulb, Borsa:2020yxh}.

In addition to the counterterms for the real emission parts of the calculation, the handling of the divergences associated with the virtual diagrams and the integrated dipoles requires the use of a regularization method. In the dimensional regularization scheme the number of dimensions is set to $d=4-2\epsilon$, and then those divergences appear as poles in $\epsilon$. Problems arise with the $\gamma^5$ matrix and the $\epsilon^{\mu \nu \rho \sigma}$ tensor since they are only properly defined in the four-dimensional spacetime. A consistent way to treat $\gamma^5$ and the Levi-Civita tensor in $d$ dimensions is the 't Hooft, Veltman, Breitenlohner, and Maison (HVBM) scheme \cite{THOOFT1972189,Breitenlohner:1977hr}, which splits the $d$-dimensional Minkowski space into the usual four-dimensional one and a $(d - 4)$-dimensional subspace where, for instance, the $(d - 4)$-dimensional part of the $\gamma^{\mu}$ matrices, represented as $\hat{\gamma}^{\mu}$, commutes with the strictly four-dimensional $\gamma^5$. Dealing consistently with $\gamma^5$ in $d$-dimensional spacetime is necessary in both polarized and electroweak DIS, since the matrix appears both in the polarization projectors and axial couplings.

\subsection{Electroweak couplings and treatment of $\gamma^5$}\label{sec_gamma5}

Due to the fact that the electroweak vertices involve the appearance of terms $\gamma^{\mu}\gamma^5$, the utilization of the HVBM scheme brings additional issues to the calculations of electroweak processes. Since the $(d-4)$-dimensional part of the $\gamma^{\mu}$ in the vertex commutes with $\gamma^5$, the location of the projector affects the $(d-4)$-dimensional part of the vertex. Thus, in order to avoid some of the issues arising in $d$ dimensions, the definition of the vertex requires the symmetrization \cite{Larin:1993tq,Korner:1989is, Buras:1989xd}

\begin{equation}
\begin{split}
 \gamma^{\mu}\gamma^5 \ \rightarrow \ \frac{1}{2}(\gamma^{\mu}\gamma^5-\gamma^5\gamma^{\mu}) = \tilde{\gamma}^{\mu}\gamma^5, 
\end{split}
\end{equation}

\noindent where $\tilde{\gamma}^{\mu}$ is the four-dimensional part of the $\gamma^{\mu}$. With this prescription, the electroweak vertex between two flavors of quarks or leptons $f_1$ and $f_2$ can be consistently expressed in terms of the vector and axial parts as

\begin{equation}
 -i e\, \gamma^{\mu}\left( C_V^{f_1 f_2} + C_A^{f_1 f_2} \gamma^5 \right) \ \rightarrow \ -i e\,\left( C_V^{f_1 f_2} \gamma^{\mu} + C_A^{f_1 f_2}\, \tilde{\gamma}^{\mu} \gamma^5 \right). 
\label{eq_vertex} 
\end{equation}


\noindent The values of the vector and axial couplings $C_V^{f_1 f_2}$ and $C_A^{f_1 f_2}$ depend on the exchanged boson and are given by the following expressions:

\begin{equation}
\begin{tabular}{l c c}
    $W^-$: & $C_V^{f_1 f_2}=\frac{1}{2\sqrt{2} \sin\theta_W}(\tau_+)_{f1 f2} \ V_{f1 f2}$, & $C_A^{f_1 f_2}=C_V^{f_1 f_2}$,\\
    $W^+$: & $C_V^{f_1 f_2}=\frac{1}{2\sqrt{2} \sin\theta_W}(\tau_-)_{f1 f2} \ V^{\dagger}_{f1 f2}$, & $C_A^{f_1 f_2}=C_V^{f_1 f_2}$,\\
    $Z$: & $C_V^{f_1 f_2}=\frac{1}{2\sin(2\theta_W)}(\tau_3)_{f1 f2}-\delta_{f1 f2}\,e_{f1}\, \tan\theta_W$, & $C_A^{f_1 f_2}=\frac{1}{2\sin(2\theta_W)}(\tau_3)_{f1 f2}$,\\
    $\gamma$: & $C_V^{f_1 f_2}= \delta_{f_1 f_2} \ e_{f1}$, & $C_A^{f_1 f_2}=0$.
\end{tabular}
\end{equation}

\noindent Here $\theta_W$ is the electroweak mixing angle, $\tau_{\pm}=(\tau_1 \pm i\tau_2)/2$ and $\tau_3$ are the weak isospin Pauli matrices, $V$ is the Cabibbo-Kobayashi-Maskawa (CKM) mixing matrix, and $e_f$ is the electric charge of the corresponding quark/lepton ($e_f = 2/3$ for $u$, $c$, $t$, $e_f = -1/3$ for $d$, $s$, $b$, and $e_f=-1$ for $e$, $\mu$ and $\tau$).

When working in the HVBM scheme, the presence of $\gamma^5$ in both the symmetrized vertex and the chirality projectors over initial-state polarized particles leads to so called ``evanescent'' terms of order $\mathcal{O}(d-4)$, which in combination with divergent terms can result in spurious finite contributions. Infrared anomalous terms cancel out between virtual and real contributions \cite{Korner:1985uj}, but in the case of the ultraviolet divergencies appropriate additional finite renormalization terms may be required to maintain the correct form of the chiral Ward identities \cite{Larin:1993tq,Ahmed:2021spj}. In this case (up to order $\alpha_s^2$), the appropriate counterterm to be subtracted is simply given by

\begin{equation}
\left( \Delta \right) C_T = \frac{\alpha_s}{2 \pi}2 C_F \left( \Delta \right) \hat{\sigma}^A_{LO},
\end{equation}

\noindent where $(\Delta) \hat{\sigma}^A_{LO}$ are the contributions to the (polarized) unpolarized leading order partonic cross section with at least one axial coupling between the weak boson and the quarks.

For the polarized case, additional finite subtraction terms are usually added within the mass factorization (in addition to those associated to the $\overline{\mathrm{MS}}$ factorization scheme) in order to enforce helicity conservation in quark lines (due to the use, by convention, of four-dimensional kernels instead of $d$-dimensional ones), as pointed out in our previous paper \cite{Borsa:2020yxh}.

We note that there are multiple ways of symmetrizing the vertex in Eq.~(\ref{eq_vertex}). For example, when working with the left and right pieces of the coupling, by performing the modification in the projectors as $\gamma^{\mu}\frac{1}{2}(1\pm\gamma^5) \rightarrow \frac{1}{4}(1\mp\gamma^5)\gamma^{\mu}(1\pm\gamma^5)$. We have checked that those choices lead to the same results. The key point is to remove the $(d-4)$-dimensional part of the $\gamma^{\mu}$ when accompanied by a $\gamma^5$ in order to avoid the appearance of spurious finite terms originating from UV-divergent pieces.

\subsection{New terms in electroweak boson exchange DIS}\label{sec_newterms}

Compared to purely photonic DIS, the inclusion of processes mediated by axial-vector bosons requires, in principle,  the calculation of additional terms proportional to the axial coupling of those bosons. It should be noted, however, that since the calculation of the matrix elements for quark-initiated processes, both for the polarized cross sections and for the axial-vector interchange cross sections, involves the presence of $\gamma^5$ in the fermionic traces, many of the new contributions associated to the axial part of the electroweak coupling can be inferred from the polarized and unpolarized matrix elements already calculated in \cite{Borsa:2020yxh} for photon exchange. For that purpose it is useful to separate the partonic cross sections into a parity-violating (PV) and a non-parity-violating (NPV) piece, as  

\begin{equation}
    \hat{\sigma}_{q}=\hat{\sigma}^{PV}_{q}+\hat{\sigma}^{NPV}_{q},
\end{equation}

\noindent where the subscript $q$ is used to indicate the contribution from quark-initiated processes. For unpolarized (polarized) DIS, the processes $q+W/Z\rightarrow q$, $q+W/Z\rightarrow q+g$ and $q+W/Z\rightarrow q+g+g$ contribute to the parity-violating part when an odd (even) number of $\gamma^5$ matrices are present in the partonic trace, while the contribution to the non-parity-violating part comes from traces with an even (odd) number of $\gamma^5$'s. It is then possible to see that, up to order $\alpha_s^2$  

\begin{equation}
\begin{split}
    \hat{\sigma}^{PV}_{q}&=\Delta\hat{\sigma}^{NPV}_{q},\\
    \Delta\hat{\sigma}^{PV}_{q}&=\hat{\sigma}^{NPV}_{q}.
\label{eq_cross_q}
\end{split}
\end{equation}

\noindent These relations are trivial for the contributions from real diagrams, since within the dipole subtraction formalism they are dealt with in four dimensions and one can anticommute $\gamma^5$ freely in the trace. Care should be taken, however, when dealing with virtual diagrams, where dimensional regularization is needed. The anticommutation of the $\gamma^5$ is only valid after performing the additional finite renormalization \cite{Larin:1991tj, Larin:1993tq}. We wish to stress that although we did calculate the new parity-violating pieces from scratch, the relations from Eq.~(\ref{eq_cross_q}) were useful to ease the implementation in the Monte Carlo code.

Analogously, the parity-violating contributions to unpolarized DIS arising from the gluon-initiated processes $g+W/Z\rightarrow q+\bar{q}$ and $g+W/Z\rightarrow q+\bar{q}+g$ can be obtained as

\begin{equation}\label{eq_cross_g}
    \hat{\sigma}^{PV}_{g}=-\Delta\hat{\sigma}^{NPV}_{q},
\end{equation}

\noindent where the minus sign arises from the crossing of the initial quark into the final state. Note that a similar relation does not hold for polarized  DIS since in this case there is no helicity projector in the quark trace. That being said, parity-violating terms in processes with an initial gluon actually vanish after integration due to charge conjugation arguments since this contribution is antisymmetric under the exchange of a parton $a$ with an antiparton $\bar{a}$ ($\hat{\sigma}_a = -\hat{\sigma}_{\bar{a}}$). These terms are only relevant if, e.g., the charge of the final jets is observed.

Special attention should be given to the contributions arising from the processes $q+W/Z\rightarrow q+q'+\bar{q'}$, since they can involve more than one partonic trace. Depending on the trace structure of the matrix element considered, relations similar to those of Eqs. \ref{eq_cross_q} and \ref{eq_cross_g} may be used. The other non-trivial cases are the new anomalous diagrams involving quark triangles, which only contribute to $Z$-boson exchange. These cancel out if the two members of each weak isospin doublet are considered.

\section{Results of Polarized NLO Dijet Production}\label{sec:dijets}

\subsection{Dijet production in neutral current DIS}\label{sec:nc}

In this section we present our results for polarized inclusive dijet production at NLO in NC DIS in the Breit frame. Since the structure of higher order corrections and the scale dependence of the cross section were already analyzed in \cite{Borsa:2020yxh}, for the case of virtual photon interchange, in the present work we focus primarily on the effects of the inclusion of the $Z$ boson contribution at NLO. As in our previous publication, we concentrate on the EIC kinematics, considering electron-proton collisions with beam energies of $E_{e}=18$ GeV and $E_{p}=275$ GeV, and we reconstruct the jets with the anti-$k_{T}$ algorithm and $E$-scheme recombination ($R=1$). The renormalization and factorization scales are fixed at central values of $\mu_{F}^{2}=\mu_{R}^{2}=\frac{1}{2}(Q^{2}+\langle p^{B}_{T}\rangle^{2})\equiv\mu_{0}^2$, with $\alpha_s$ evaluated at NLO accuracy with $\alpha_s(M_z)=0.118$, and we require that the pair of leading jets satisfy the following kinematical cuts:

\begin{equation}
\begin{tabular}{ c }
    $p^{B}_{T,1}>5\ \mathrm{GeV}$,\\
    $p^{B}_{T,2}>4\ \mathrm{GeV}$,\\
    $|\eta^{L}|<3.5$,
\end{tabular}
\end{equation}

\noindent where the $p_T$ and $\eta$ cuts are imposed in the Breit and laboratory frames, respectively. The lepton kinematics is restricted by

\begin{equation}
\begin{tabular}{ c }
    $0.2<y<0.6$,\\
    $25\, \mathrm{GeV}^{2}<Q^{2}<2500 \, \mathrm{GeV}^{2} $.
\end{tabular}
\end{equation}

\begin{figure}
 \epsfig{figure= 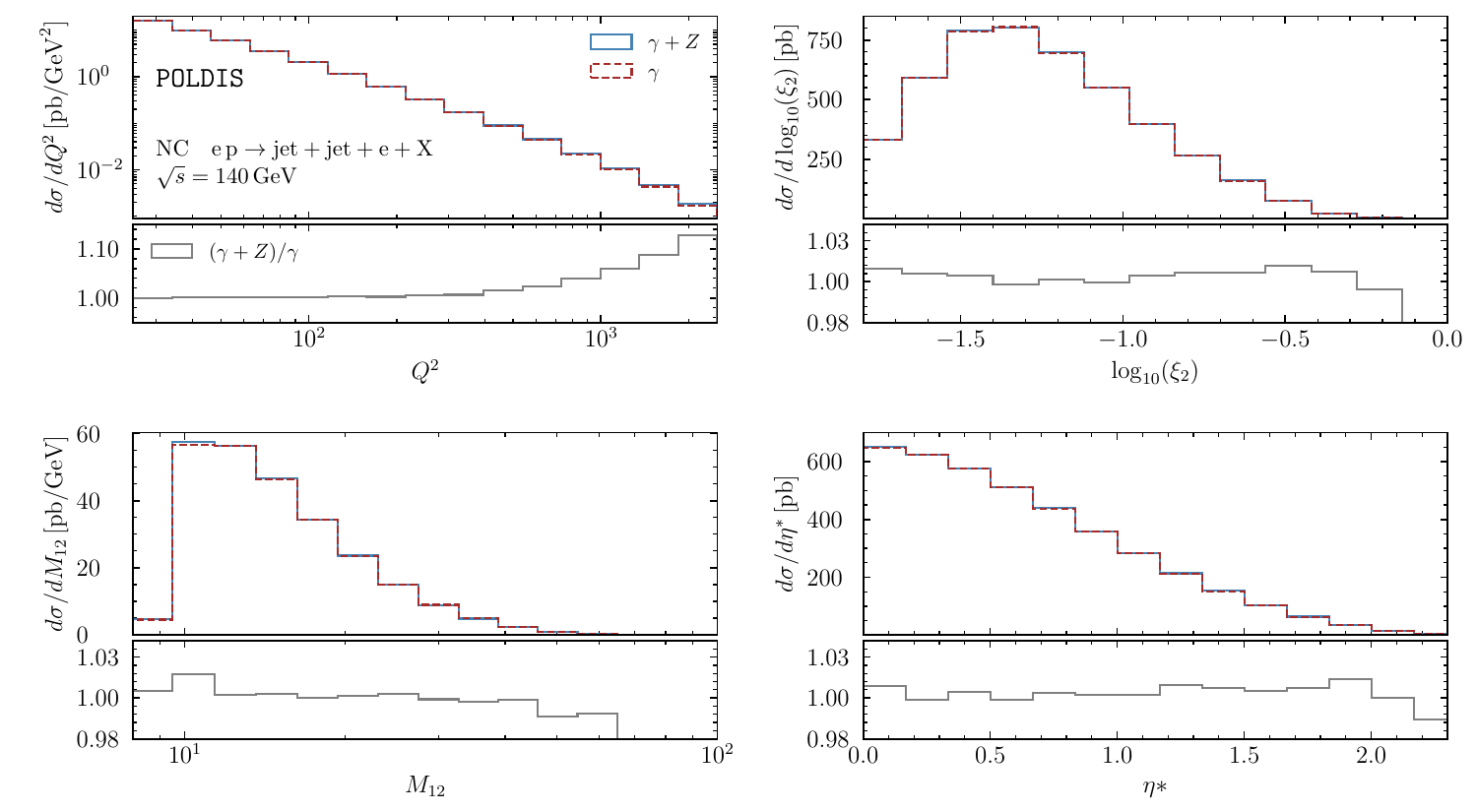, width=0.98\textwidth}
  \caption{Inclusive dijet production distributions as a function of the variables $Q^2$, $\log_{10}(\xi_2)$, $M_{12}$ and $\eta^*$, for DIS mediated by pure photon (red) or full NC $Z/\gamma$ (blue). The lower boxes show the ratio between the two cross sections.}\label{fig_dist_nopol}
\end{figure}

\begin{figure}
 \epsfig{figure= 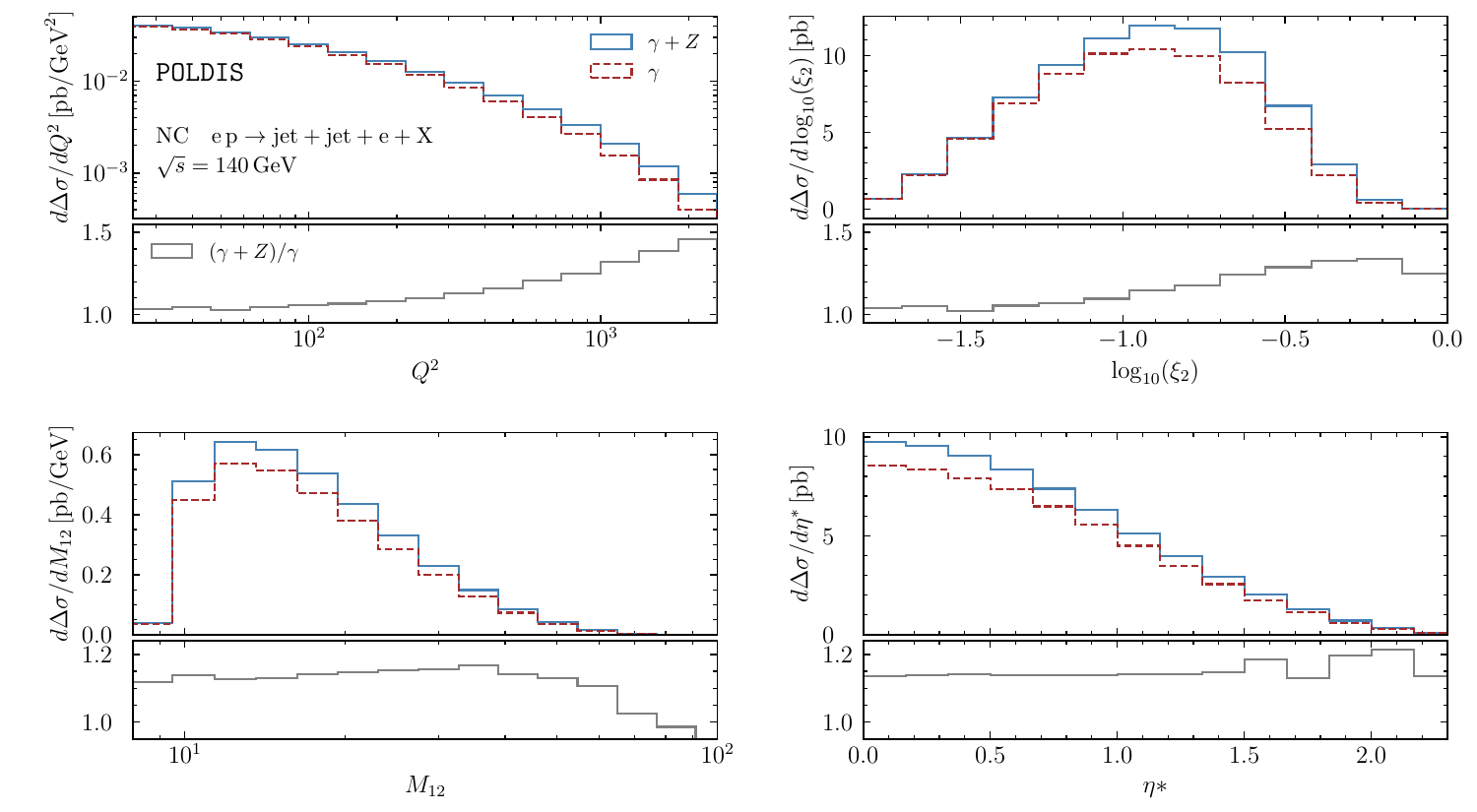, width=0.98\textwidth}
  \caption{Same as Fig. \ref{fig_dist_nopol}, but for the polarized case. }\label{fig_dist_pol}
\end{figure}


For the $Z$ boson we use a mass $M_z = 91.1876$ GeV and a decay-width of $\Gamma_z = 2.4952$ GeV, with an electromagnetic coupling
constant $\alpha = 1/137$ and the Weinberg angle given
by $\sin^2\theta_W = 0.23122$. The parton distributions sets used were the NLOPDF4LHC15 \cite{Butterworth:2015oua} and DSSV \cite{deFlorian:2014yva,deFlorian:2019zkl} for the unpolarized and polarized cases, respectively.

We start by presenting the NLO results for the dijet production cross section in unpolarized DIS in Fig. \ref{fig_dist_nopol}, both for processes mediated exclusively by a virtual photon as well as for the full neutral current interchange ($Z/\gamma$). The cross sections are presented as distributions in the boson virtuality $Q^2$, the logarithm of the dijet momentum fraction $\log_{10}(\xi_2)$, the invariant mass of the dijet system $M_{12}$, and the pseudorapidity difference in the Breit frame $\eta^*$. The lower insets in the figure show the ratio between the $Z/\gamma$ mediated cross section and the one with pure photon exchange. As has already been noted by \cite{Mirkes:1997sp, Mirkes:1997uv}, due to the $Z$ boson propagator suppression, the contribution to the cross section from processes with a $Z$ are small for the values of $Q^2$ to be covered by the future EIC (surpassing 5\% only for $Q^2>1000$), and comes fundamentally from the $Z\gamma$ interference terms.

Fig. \ref{fig_dist_pol} presents the same distributions as Fig. \ref{fig_dist_nopol}, but for longitudinally polarized scattering. While the polarized cross section has the same propagator suppression as the unpolarized case, the corrections associated to the $Z$ boson exchange are far more pronounced for the former. As a function of $Q^2$, the increase of the cross section ranges between 5\% (low $Q^2$) and 50\% (high $Q^2$), while for the remaining distributions it is typically of order 15\%. The larger EW effects in the polarized case are associated to the fact that the contributions coming from different partonic channels can have relative signs, leading to cancellations between channels. Since those cancellations result in a suppression of the cross section at low $Q^2$, if the $Z$ boson corrections are not the same for each of the channels canceling out, the relative corrections can be enhanced. This is exactly the case for dijet production: for polarized scattering, the contribution from gluon-initiated processes is negative and becomes comparable to that of the quarks for low $Q^2$, as the emission of soft gluons becomes more prominent. Since the gluon contribution to the parity-violating piece of the cross section (related to the $g_4$ and $g_5$ structure functions) cancels after the integration over the phase space, while the quark net contribution is non-zero, the inclusion of $Z$ boson interchange leads to more sizable corrections. This difference is significant over the whole $Q^2$ range, since at high $Q^2$ the cancellation between the $u$ and $d$ flavors also becomes relevant. As mentioned, a detailed discussion of the different partonic contributions to the dijet cross section, both in the polarized and unpolarized cases, can be found in \cite{ Borsa:2020yxh}.

\begin{figure}[!tb]
 \epsfig{figure= 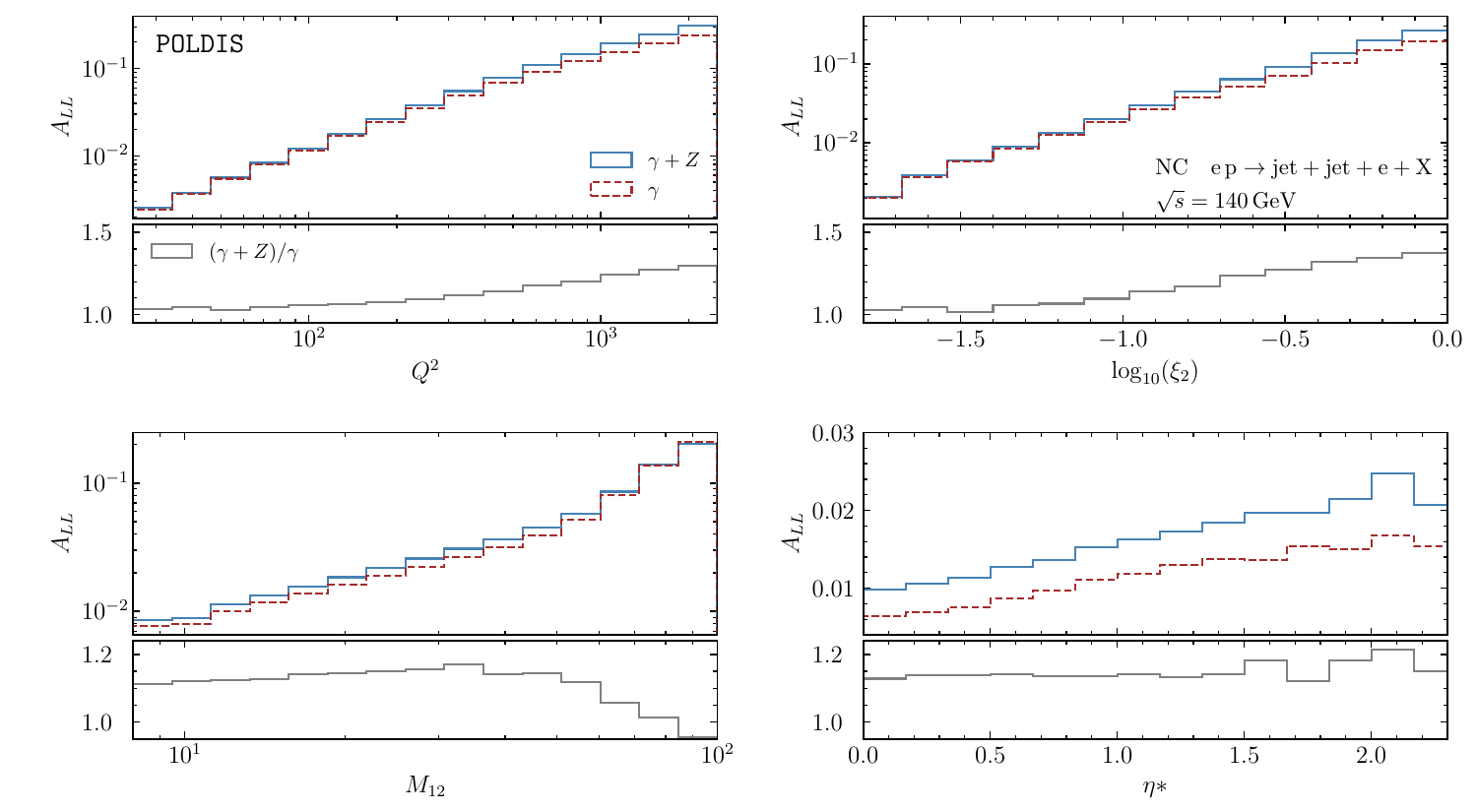, width=0.98\textwidth}
  \caption{Double spin asymmetries for dijet production, as a function of $Q^2$, $\log_{10}(\xi_2)$, $M_{12}$ and $\eta^*$, for DIS mediated by pure photon (red) or full NC $Z/\gamma$ (blue). The lower boxes show the ratio between the asymmetry for $Z/\gamma$ exchange, and that of purely photonic exchange.}\label{fig_dist_asym}
\end{figure}

The difference in the magnitude between the $Z$ boson contribution and the polarized and unpolarized distributions is reflected in the double spin asymmetries, presented in Fig. \ref{fig_dist_asym}, which are defined as the ratio of the polarized to the unpolarized cross sections $A_{LL}=\Delta \sigma / \sigma$. The asymmetry is enhanced at the high $Q^2$ region, where it increases up to 30\% with respect to the pure photon exchange. As high $Q^2$ correlates with higher values of $\xi_2$, the asymmetry also increases significantly at dijet momentum fractions closer to $1$. Conversely, in both the asymmetries in $M_{12}$ and $\eta^*$, this enhancement is more evenly distributed, with ratios of $\sim 1.13$ over most of the available range. It should be noted that, due to the origin of this behavior, the enhancement of the asymmetries is already present at the LO. Overall, due to the combination of the channel cancellations in the polarized process and the nature of the NC contributions, the double spin asymmetries are noticeably increased even at the relatively small values of $Q$ (compared to $M_Z$) to be reached at the EIC, making it an extremely useful tool to disentangle the contribution from the different partonic channels, particularly the less constrained sea-quark distributions. 

\subsection{Dijet production in charged current DIS}\label{sec_cc}

In this section we present our results for polarized inclusive dijet production at NLO in CC DIS. The cut imposed on the jets and the reconstructed lepton kinematics are the same as the ones in the previous section. For the $W$ boson we use a mass $M_W = 80.379$ GeV and a decay-width of $\Gamma_W = 2.085$ GeV. The values used for the CKM matrix are $|V_{ud}|=0.9737$, $|V_{us}|=0.2245$, $|V_{ub}|=0.00382$, $|V_{cd}|=0.2210$, $|V_{cs}|=0.987$ and $|V_{cb}|=0.041$.

\begin{figure}[!tb]
 \epsfig{figure= 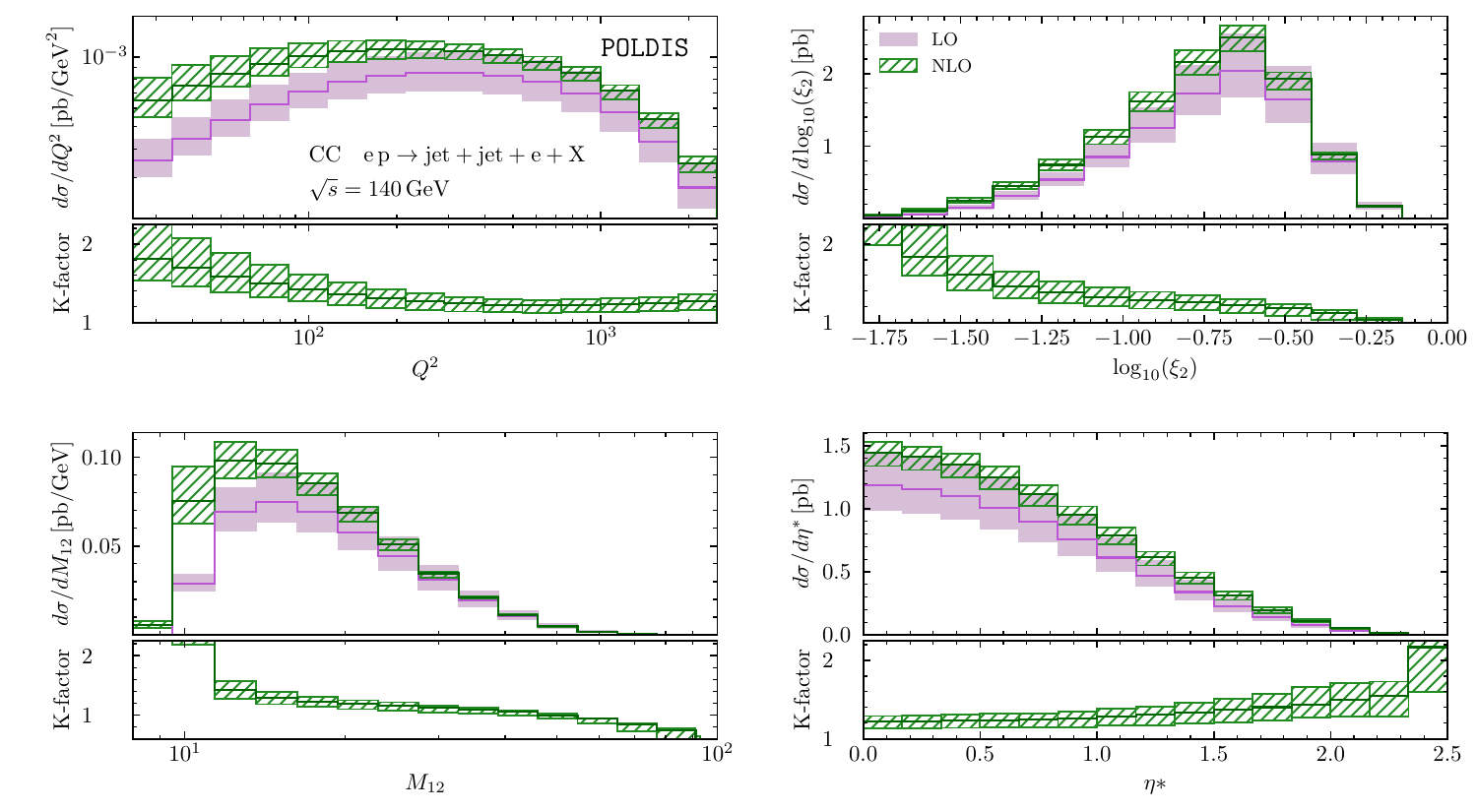, width=0.98\textwidth}
  \caption{Cross section for dijet production in unpolarized, charged current electron-proton  DIS, as a function of $Q^2$, $\log_{10}(\xi_2)$, $M_{12}$ and $\eta^*$, at Leading (pink) and Next-to-Leading Order (green). The bands correspond to the theoretical uncertainty obtained by performing the seven-point variation of the factorization and renormalization scales.  The lower boxes show the K-factor, i.e, the ratio of the NLO cross section to the LO one.}\label{fig_dist_w_nopol}
\end{figure}

We start by analyzing unpolarized CC dijet production in the EIC kinematics. In Fig. \ref{fig_dist_w_nopol} we present the unpolarized LO and NLO cross sections as a function of the $W$ boson virtuality $Q^2$, $\log_{10}(\xi_2)$, the invariant mass $M_{12}$ and the pseudorapidity difference $\eta^*$. The bands shown correspond to the estimation of the theoretical uncertainty, obtained by performing the seven-point variation of the factorization and renormalization scales as $\mu_R,\mu_F = [1/2,2]\, \mu_{0}$ (with the additional constraint $1/2\leq\mu_{F}/\mu_{R}\leq2$). The lower box in each distribution shows the corresponding K-factor, defined as the ratio to the LO cross section $\sigma_{LO}$, in order to quantify the effect of the higher order corrections. To that end, the same NLO sets of PDF were also used for the calculation of the LO distributions.

As expected, the distributions shown in Fig. \ref{fig_dist_w_nopol} are highly suppressed compared to the NC ones, especially at low $Q^2$, due to the massive boson propagator. The propagator suppression also accounts for the shift in the $\xi_2$ distributions to higher momentum fractions, since $Q^2$ and $\xi_2$ are correlated. This means that in this case the kinematics limit the probing of small momentum fractions of the initial parton. In terms of the structure of higher order corrections, while the NLO distributions show a general reduction of the scale dependence going to NLO, the corrections are still sizable. As mentioned in \cite{Borsa:2020yxh}, the minimum-$p_T$ cuts imposed on the dijet systems imply that the regions $M_{12}>10$ GeV and $\xi_2\gtrsim5\times10^{-3}$ are forbidden at LO and become available only from NLO, making the calculation for those regions effectively LO, and leading to perturbative instabilities. Since $Q^2$ correlates with $M_{12}$, the slow perturbative convergence is also observed for low-$Q^2$ values.

\begin{figure}[!tb]
 \epsfig{figure= 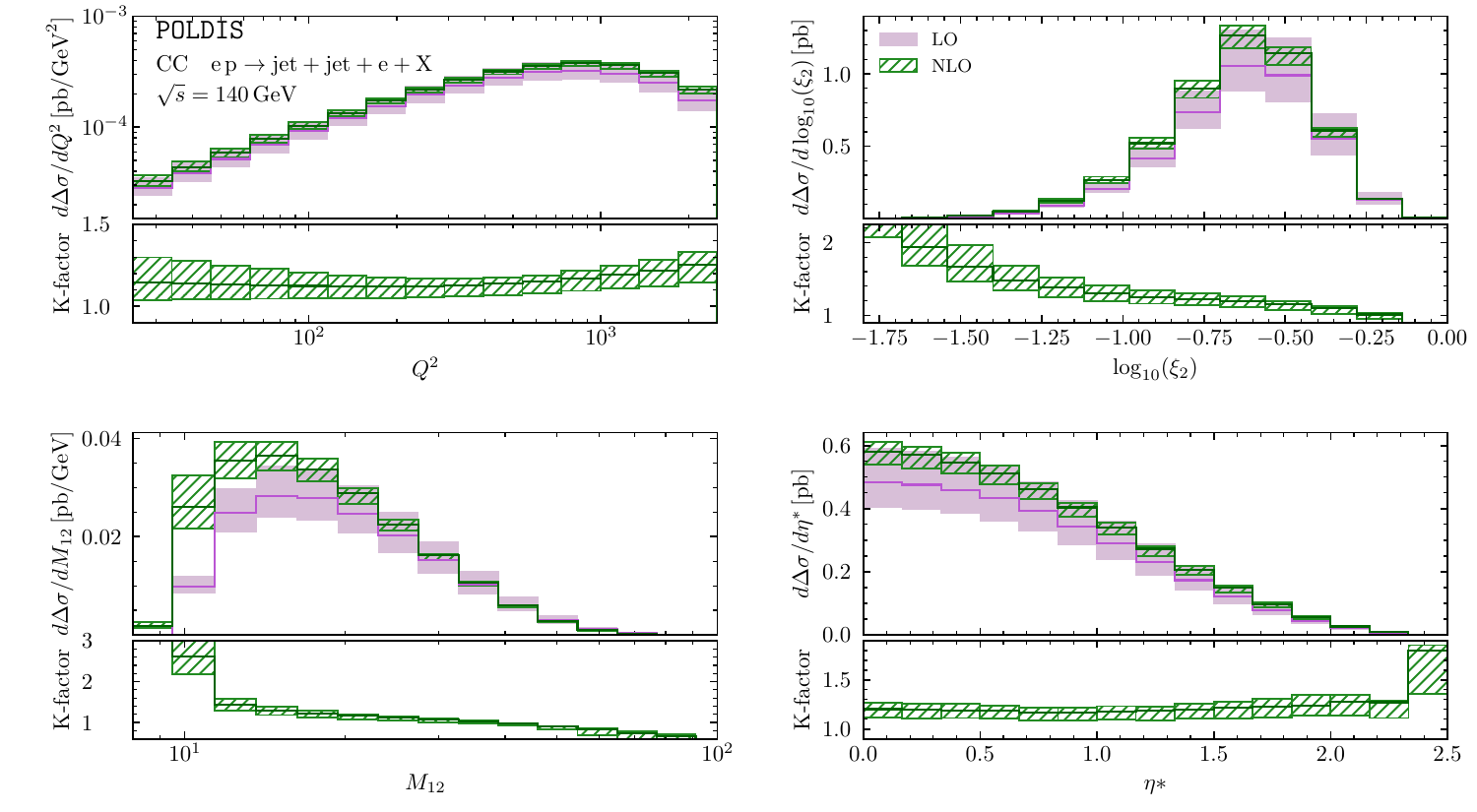, width=0.98\textwidth}
  \caption{Same as in Fig. \ref{fig_dist_w_nopol}, but for polarized electron-proton scattering.}\label{fig_dist_w_pol}
\end{figure}

In Fig. \ref{fig_dist_w_pol} we present the polarized CC dijet distributions. It is interesting to notice that, as opposed to the NC case, the distributions for polarized DIS are not as suppressed compared with those for unpolarized DIS (differing only up to an order of magnitude at low-$Q^2$). Due to the nature of the $W$ boson, having equal vector and axial couplings, the parity-violating part of the cross section is not suppressed as in the NC case. Since this contribution stems only from initial-quark processes, the initial-gluon channel becomes overall less relevant, thus reducing the cancellations between channels in the polarized cross section. When comparing to the unpolarized case, this effect is further amplified by a suppression of the gluon helicity distribution at lower momentum fractions. The different low-$Q^2$ behavior of the polarized cross section and K-factors is then a consequence of the smaller relevance of the gluonic contribution. 

\begin{figure}[!tb]
 \epsfig{figure= 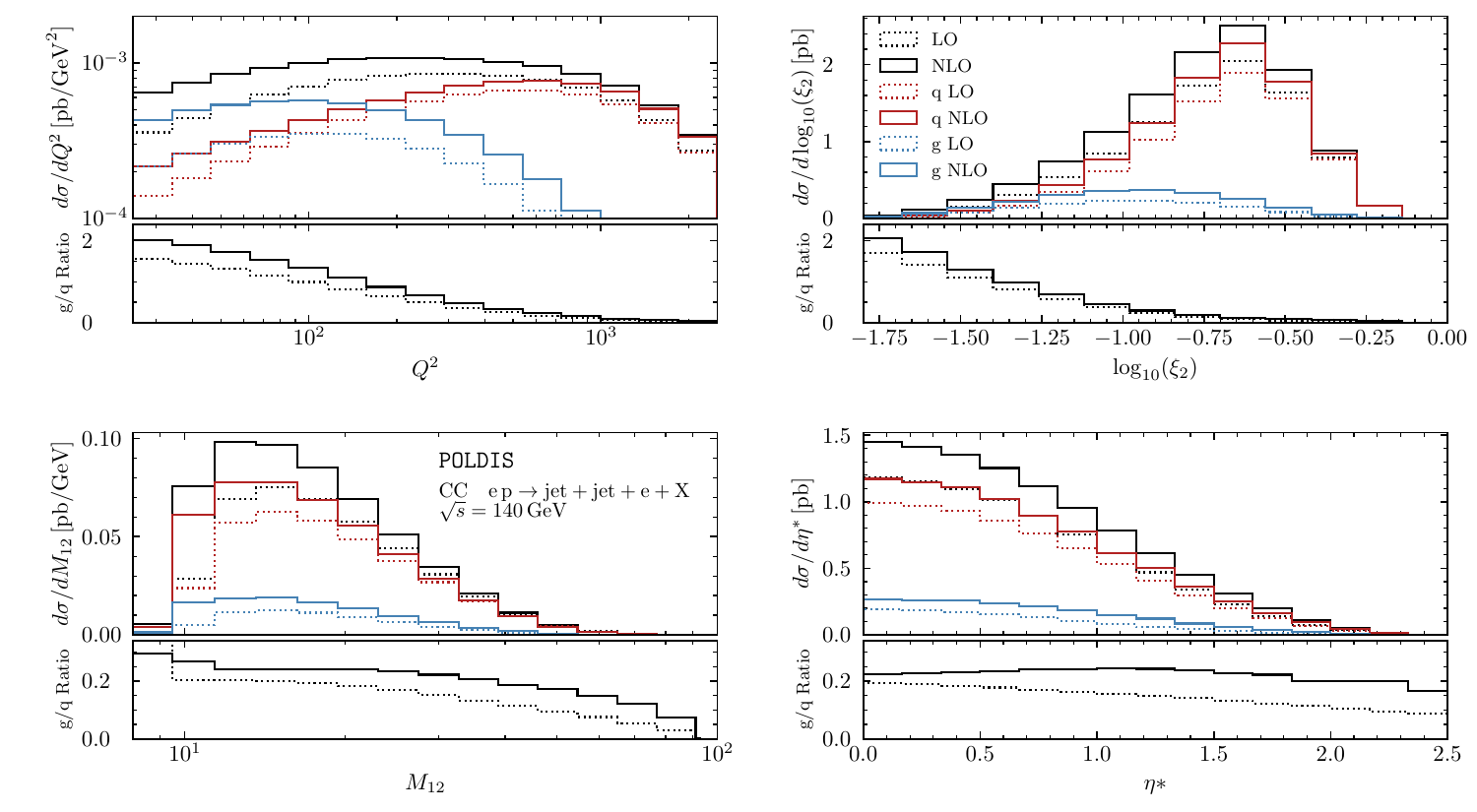, width=0.98\textwidth}
  \caption{Same distributions as in Fig. \ref{fig_dist_w_nopol}, but separating the contributions of quark and gluon-initiated processes. The lower insets show the ratio between the two partonic channels}\label{fig_dist_w_nopol_canales}
\end{figure}
\begin{figure}[!tb]
 \epsfig{figure= 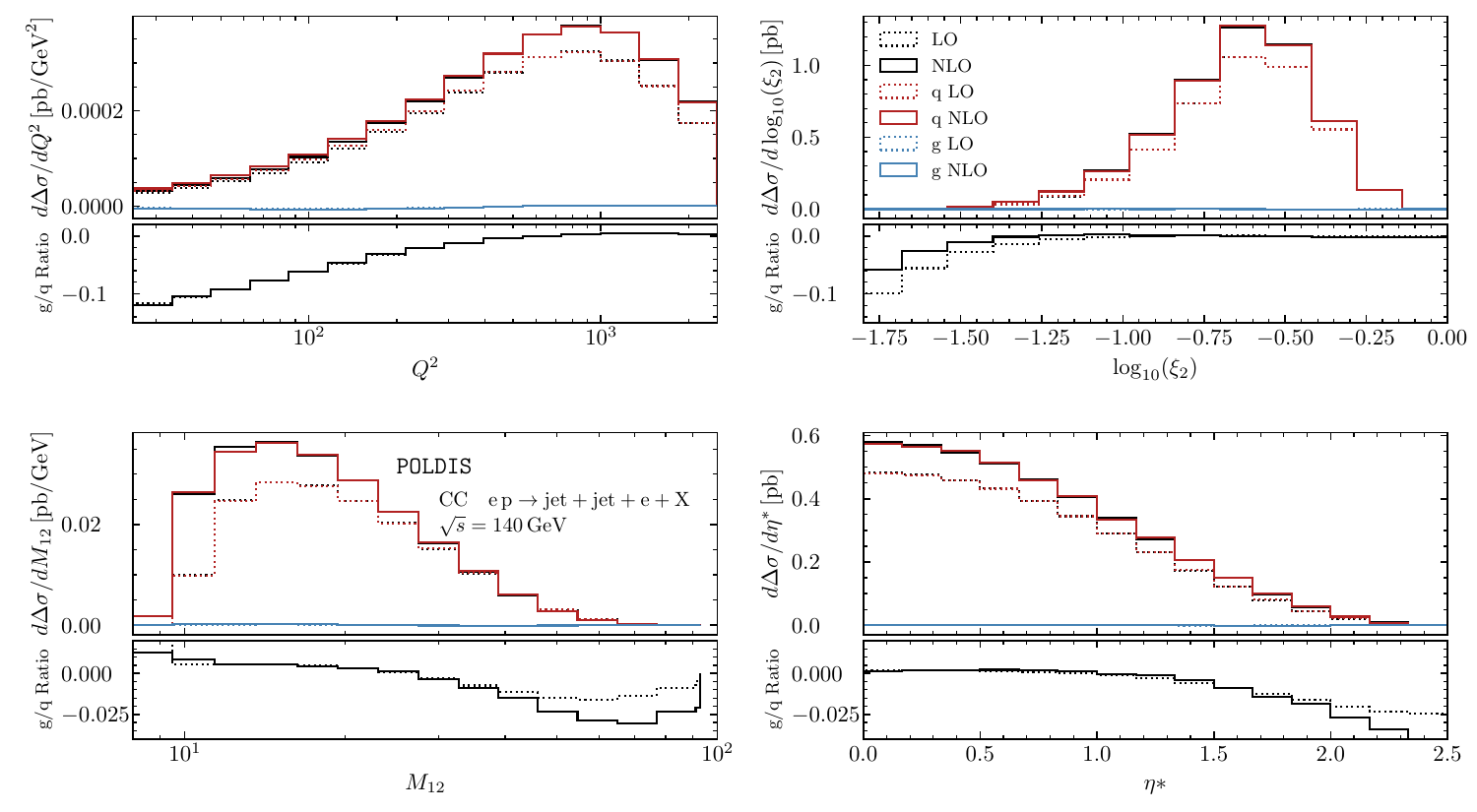, width=0.98\textwidth}
  \caption{Same as in Fig. \ref{fig_dist_w_nopol_canales}, but for polarized electron-proton scattering.}\label{fig_dist_w_pol_canales}
\end{figure}

To highlight this point, Figs. \ref{fig_dist_w_nopol_canales} and \ref{fig_dist_w_pol_canales} present the same distributions as Figs. \ref{fig_dist_w_nopol} and \ref{fig_dist_w_pol}, but distinguishing the quark and gluon contributions to the cross section. The lower boxes show the ratios between the gluon-initiated pieces and the quark-initiated ones. While for the unpolarized distribution, the low-$Q^2$ cross section is mainly driven by the gluon channel (reaching ratios of up to 2), the aforementioned suppression in the polarized case results in a significantly smaller cross section in this region. This general reduction of the relevance of the gluon-initiated process can also be observed in the ratios for the other distributions.


\begin{figure}
 \epsfig{figure= 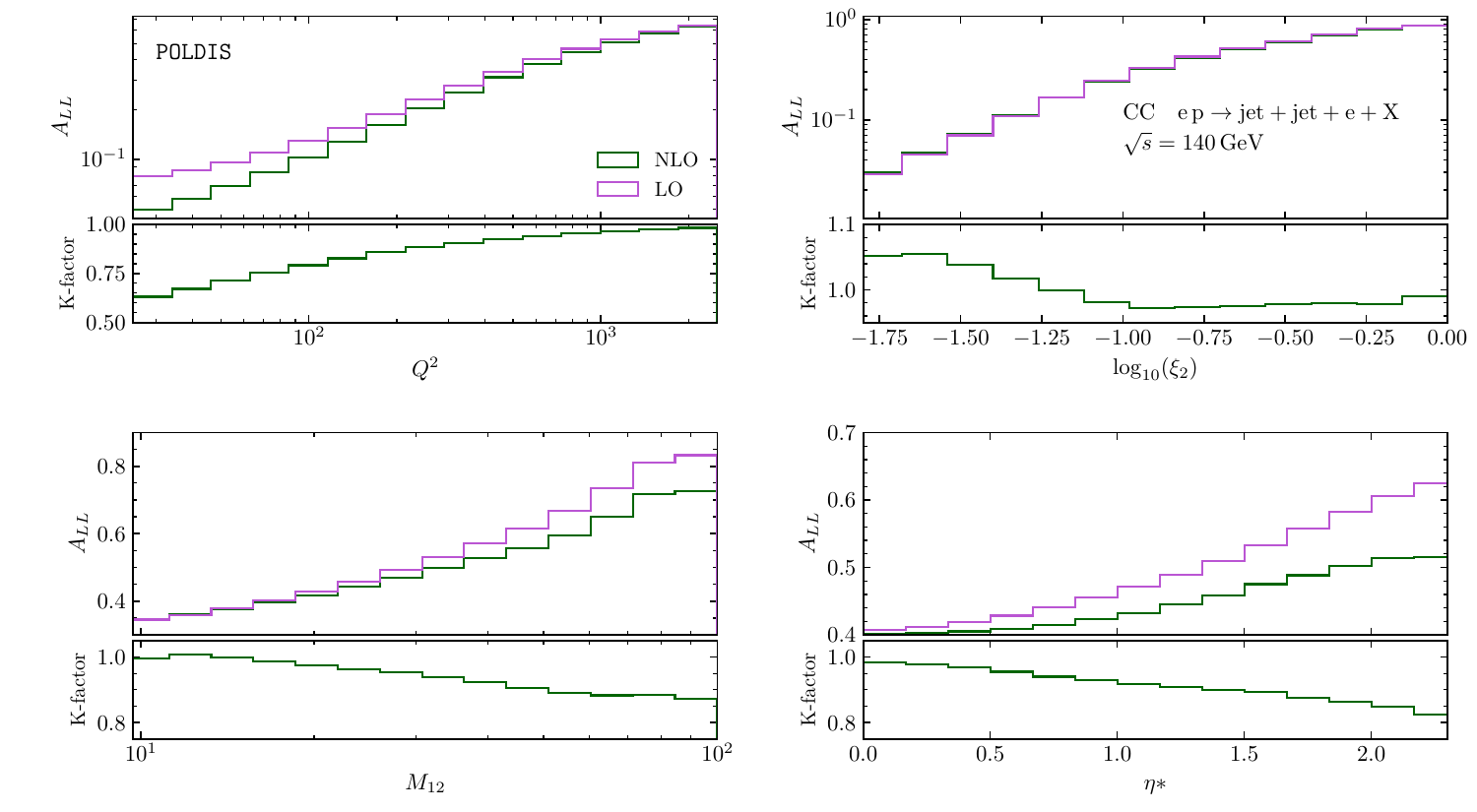, width=0.98\textwidth}
  \caption{Double spin asymmetries as distributions in $Q^2$, $\log_{10}(\xi_2)$, $M_{12}$ and $\eta^*$ for CC electron-proton scattering.}\label{fig_dist_asym_w}
\end{figure}

The reduction of cancellations between channels in the polarized cross section leads, in turn, to higher values of the double spin asymmetries with CC, defined as in section \ref{sec:nc}, which are presented in Fig. \ref{fig_dist_asym_w}. This effect also results in milder K-factors of the asymmetry (compared to the photon-exchange case in \cite{ Borsa:2020yxh}), with values of $\sim 0.7$ at low-$Q^2$, that then approach unity at higher boson invariant mass. Note that, in spite of the small value of the cross section, the asymmetries for charged current DIS are typically of order $\sim 0.8$ for the relevant high-$Q^2$ region, providing an additional valuable constraint to the helicity parton distribution functions.

\section{Conclusions}\label{sec:conclusion}

In this paper we presented the fully exclusive NLO calculation of dijet production in polarized DIS, considering the exchange of both neutral and charged currents. The calculation was performed with the dipole subtraction method, which was extended to allow for longitudinal polarization of initial partons. In particular, we analyzed the production of dijets in the Breit-frame with EIC kinematics. The cross sections were studied as distributions in the virtuality $Q^2$, invariant jet mass $M_{12}$, the pseudorapidity difference $\eta^*$, and the dijet momentum fraction $\xi_2$. For NC DIS, the contribution of the $Z$ boson to the unpolarized cross section is, as expected, very small overall, with the exception of the highest $Q^2$ bins for which the enhancement is of order 10\%. In the polarized case, however, due to the cancellations between partonic channels, the effect is not only non-negligible at low $Q^2$, but can reach 50\% for the high-$Q^2$ region, leading to an increase of the double spin asymmetry. In the case of the distributions for CC DIS, while an overall reduction in the scale dependence is observed for the higher order terms, the corrections are still sizable. In terms of the experimental observables, higher asymmetries are obtained compared with the NC case, mainly due to a strong suppression of the gluon channel (which is negative for polarized DIS), reaching values of $\sim 0.8$ at higher $Q^2$.

The results presented in this paper highlight the potential of the electroweak boson exchange processes to further improve our knowledge on polarized parton distributions, providing complementary constraints to determine the flavor decomposition of the proton spin, as well as the relevance that higher order QCD corrections will have in the precise description of the jet observables to be measured in the future EIC. Our dijet results will also be a fundamental step towards the obtention of the fully exclusive NNLO calculation for jet production in DIS with electroweak boson exchange.

\acknowledgments

I.B. wishes to thank the University of Tübingen for its hospitality during the last stages of this project. 
This work was partially supported by CONICET and ANPCyT.

\bibliography{refs}
\clearpage

\end{document}